\documentclass[fleqn,usenatbib]{mnras}

\usepackage{newtxtext,newtxmath}

\usepackage[T1]{fontenc}

\DeclareRobustCommand{\VAN}[3]{#2}
\let\VANthebibliography\thebibliography
\def\thebibliography{\DeclareRobustCommand{\VAN}[3]{##3}\VANthebibliography}

\usepackage{pdflscape}
\usepackage{graphicx}	
\usepackage{amsmath}	

\title[Reading Between the Rainbows]{Reading Between the Rainbows: Comparative Exoplanet Characterisation through Molecule Agnostic Spectral Clustering }

\author[I.A. Guez]{
I.A.Guez,$^{1}$\thanks{E-mail: iacg1@st-andrews.ac.uk}
Mark Claire,$^{2}$
\\
$^{1}$School of Earth and Environmental Sciences, University of Saint Andrews,
College Gate, KY16 9AJ, Scotland\\
$^{2}$Blue Marble Space Institute of Science, 600 1st Avenue, 1st Floor, Seattle, Washington 98104\\
}

\date{Manuscript Oct 2024}

\pubyear{\the\year{2024}}

\begin{document}
\label{firstpage}
\pagerange{\pageref{firstpage}--\pageref{lastpage}}
\maketitle

\begin{abstract}
Rocky exoplanets are faint and difficult to observe due to their small size and low brightness compared to their host star. Despite this, the James Webb Space Telescope (JWST) has allowed us new methods and opportunities to study them. Accurately characterising exoplanet atmospheres could offer insights not only into the planetary demographics of rocky worlds in the universe, but also how our own Earth compares. 
Previous work simulating the observational spectra of planets with various models of atmospheric composition has constrained conditions under which specific molecules would be observable. We seek a different approach that does not depend on specific molecular features, but rather on correlating "agnostic" spectral characteristics to each other.  This study uses 42 synthetic transit transmission spectra in a range matching that of JWST’s MIRI instrument. Spectra were de-noised, normalised and split into bands for which the enclosed areas were calculated. Using the HDBSCAN clustering algorithm, separation patterns and clusters were found across these bands, which correlated to various \textit{a priori} planet characteristics, such as relative gas concentration. Using 0.7 $\mu$m bands between 7 and 12 $\mu$m, we are able to separate reducing vs. oxic atmospheres and constrain CO$_2$ and O$_2$ mixing ratios accurate to within two orders of magnitude in an unknown sample.
Through this, we have demonstrated that it is possible to differentiate and analyse exoplanet atmosphere type from agnostic interrogation of their transit spectra with a method we call ``Molecule Agnostic Spectral Clustering''.

\end{abstract}



\section{Introduction}

Terrestrial exoplanets are a mysterious feature of our universe and the subjects of much curiosity. Our own planet being of the rocky sort, the study of these rocky worlds outside our solar system is a window into Earth’s own extended family. To date, several missions have aimed to create a better census of these planets such as the Transiting Exoplanet Survey Satellite (TESS, launched in 2018) \citep{RickerTESS2014} which tiles the sky in search of transiting exoplanets, and Kepler (launched in 2009) \citep{Brown_2011} whose nine-year mission resulted in 2800 exoplanet discoveries. Altogether, upwards of 5750 exoplanets have been discovered  to date \citep{ExArch}.

Exoplanets, being non-light emitting objects, are difficult to observe. The most reliable way of detecting them to date has been either through their shadow cast upon their star (transit), or in their gravitational influence on their star (radial velocity). These methods, however, introduce a bias towards large planets, which cast larger shadows and have more pronounced effects on their stars. Smaller terrestrial planets, thus, are often nearly entirely engulfed in the light of their star. Because of this bias, discovering terrestrial exoplanets, let alone observing their atmospheres, presents a particular challenge. It is due to these limitations in the detection methods, that only about 100 of the exoplanets discovered are confirmed to be low-mass ($<2.5R_E$, $<10M_E$) terrestrial planets \citep{lichtenberg2024, Madhusudhan2019}.

This does not, however, mean that study is impossible. The recently launched James Webb Space Telescope (JWST) is capable of detecting infrared wavelengths with modern instruments and improved resolution over previous telescopes \citep{JWSTDocs2016}, and offers us new opportunities to study exoplanets and make strides towards characterising them. JWST observations have made it possible to observe and analyse transit transmission spectra of rocky exoplanets and begin to gain insight into their atmospheric composition \citep{Deming_2009, Lim_2023}, should the resolution be high enough. Once again, this is an easier process on giant gas planets, who have thicker- or more vertically extended- atmospheres, than on the smaller rocky exoplanets \citep{Madhusudhan2019, Bean_2017, Lustig-Yaeger_2022}.

\subsection{Agnostic versus Specific Analysis of Spectral Patterns}
Where a molecular absorption band occurs, the amount of light passing through the atmosphere decreases, causing the transit depth at that specific wavelength to grow. Since different atmospheric species absorb at different wavelengths, at first glance this should allow us to read the spectrum like a barcode. However, these absorption lines can reoccur depending on the species, and can overlap with other species or with atmospheric and environmental features, obscuring this barcode. Prior work on the characterization of terrestrial exoplanet atmospheres \citep[e.g.,][]{Zieba2023-cf, Lustig-Yaeger_2019, Meadows_2023} has focused on exploring the detectability of specific molecules in a spectrum, primarily  considering signal-to-noise constraints on individual strong lines.

Here, we present an agnostic approach to spectral analysis, considering the spectrum purely numerically. The interactions between spectral features, the evidence of molecular interactions within the atmosphere in a spectrum, and even haze or cloud interference, could then act as potential determinants in our attempts to characterise exoplanetary type, if they were shown to distinctly correlate with \textit{a priori} knowledge. We compare multiple agnostic observations of our modelled planetary spectra and identify patterns which correlate with our prior knowledge of the underlying atmospheres, then demonstrate that this method enables quantitative classification of spectra beyond the presence or absence of a given species, motivating additional study of the technique we call ``Molecule Agnostic Spectral Clustering" or MASC.

\section{Methods}
\subsection{Models}
For this study, we used transit transmission spectra found in the VPL Spectral Explorer (VPLSE). This database contains the spectra featured in many of the Virtual Planetary Laboratory’s studies and their associated metadata \citep{Arney_2017, Lincowski_2018, Meadows2018, Segura2005, Schwieterman_2016}. These modeled transmission spectra are given in terms of transit depth \((\frac{R_p}{R_s})^2\), where R$_p$ is the planetary radius and R$_s$ is the stellar radius. Our curated dataset is comprised of 42 planets, which we roughly organize into eight distinct initial condition archetypes as compiled in Table \ref{tab:Symbols}. Additional detailed information on all 42 models is included as Appendix \ref{sec:Plan}. This relatively small dataset allows us the freedom to efficiently examine many different types of planets while retaining a sufficient number of points to demonstrate the feasibility of the MASC technique.

\begin{table}
 \caption{A summary of the planetary types found in the VPLSE dataset}
 \centering
 \begin{tabular}{l | c | l}
 \hline
  Name  & Symbol & Description  \\
 \hline
   $Archean$  & red X & Reducing Archean-like atmospheres\\
   $ArcheanHaze$  & red star & Hazy Archean-like atmospheres\\
   $E-type$  & navy dot  & Oxic Earth-like atmospheres \\
   $ModernEarth$  & blue cross & A model of the modern Earth \\
   $DryO_2$  & purple X & Dry oxygen atmospheres \\
   $WetO_2$  & blue star & Wet oxygen atmospheres \\
   $V-type$  & green cross & Venus-like atmospheres \\   
   $CO_2$  & yellow dot & CO$_2$ atmospheres \\
 \hline
 \end{tabular}
 \label{tab:Symbols}
 \end{table}
 
The models found in the VPLSE were computed at a much higher resolution, and with greater spectral range, than current telescopes can observe. In order to assess the realistic applications of our technique, we must attempt to match the wavelength range of a known instrument and reduce the resolution of the spectra. In order to compare spectra of planets with differing orbital geometries around stars of varying size, we normalise all spectra to their lowest transit depth value. This removes our ability to extract information about the relative height of atmospheric radiation, which we accept as a limitation of the MASC method.

\begin{figure}
 \noindent\includegraphics[width=\columnwidth]{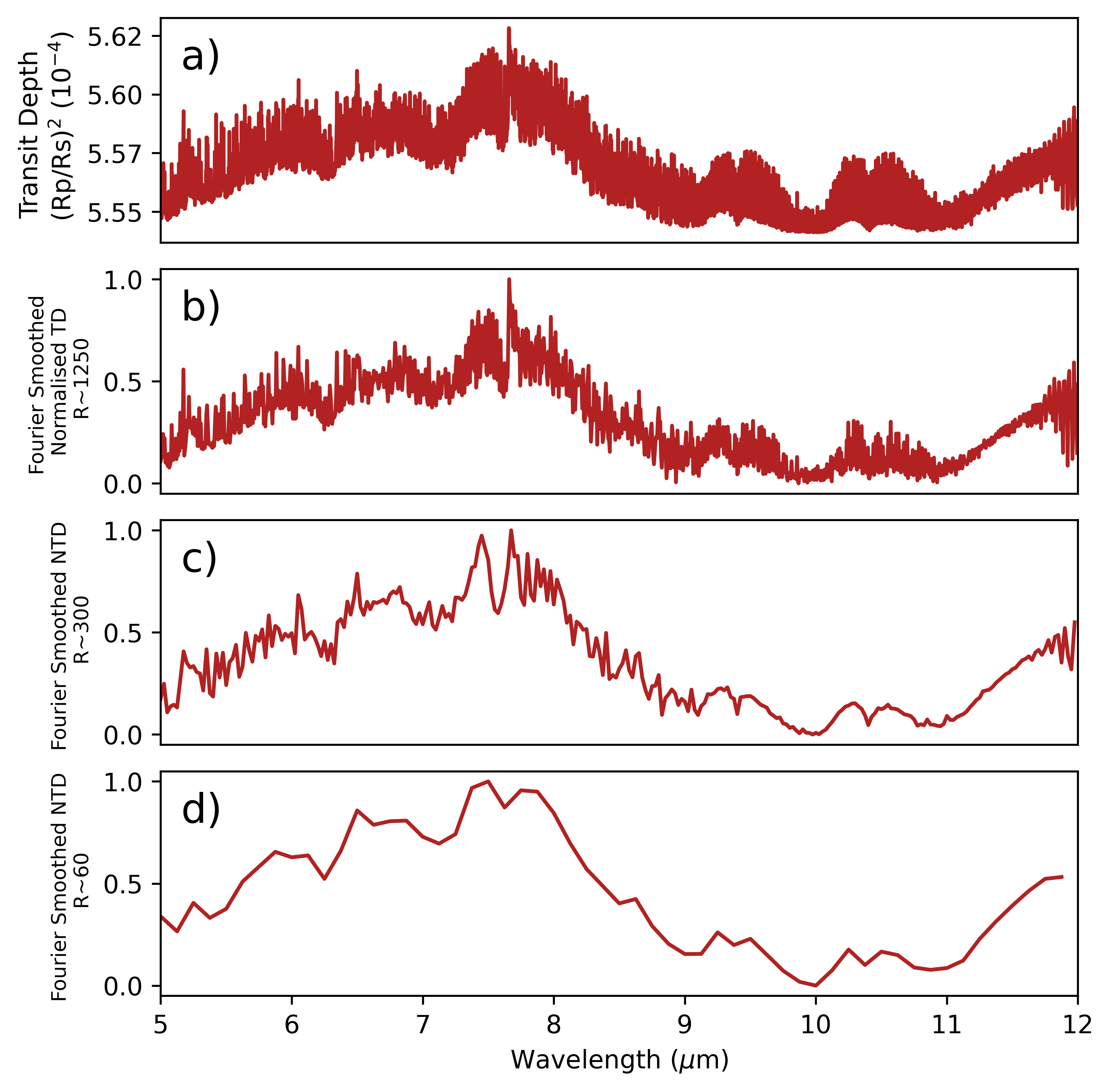}
 \caption{Reducing the resolution of the synthetic spectral data of an Archean Earth-equivalent from the VPLSE using two-way FFT smoothing. Panel a) shows the full spectrum as found in the VPLSE, b) shows R$\sim$1250 approximating MIRI MRS, c) shows R$\sim$300 d) shows R$\sim$60 approximating MIRI LRS}
 \label{fig:FFTDegrade}
\end{figure}

In Figure \ref{fig:FFTDegrade} we show the VPLSE transit transmission spectrum of a reducing Earth-like planet at 1 AU equivalent distance from AD Leo, and several normalised, reduced resolution spectra based on that data. This was achieved by performing the forward Fast Fourier Transform (FFT), reducing the number of points to 1166, 280 and 56, then performing the inverse FFT. This results in the highest frequency oscillations being excluded from the reduced plots, which have lower resolution which more closely match real observations with varying degrees of signal loss.

These values correspond to resolving powers (R) of 1250, 300 and 60, the first and third of which were chosen to conservatively simulate the specifications of JWST's Mid-Infrared Instrument (MIRI) \citep{JWSTDocs2016}. MIRI's Medium Resolution Spectrometer mode (MRS) has resolving power ranging from 1500 - 3500 over 5-12 $\mu$m \citep{JonesMIRI2023}, while MIRI's Low Resolution Spectrometer (LRS) has resolving power ranging from 40 - 160 \citep{Kendrew_2015}.  In both instrument cases, we adopted the average resolution over the range, then halved that value to account for interference and noise. The R$\sim$300 spectrum is displayed as an experimental intermediate case.

\subsection{Tools}
\label{sec:tools}
\subsubsection{Integration Bands}
\label{sec:IntBands}
\begin{figure}
\noindent\includegraphics[width=\columnwidth]{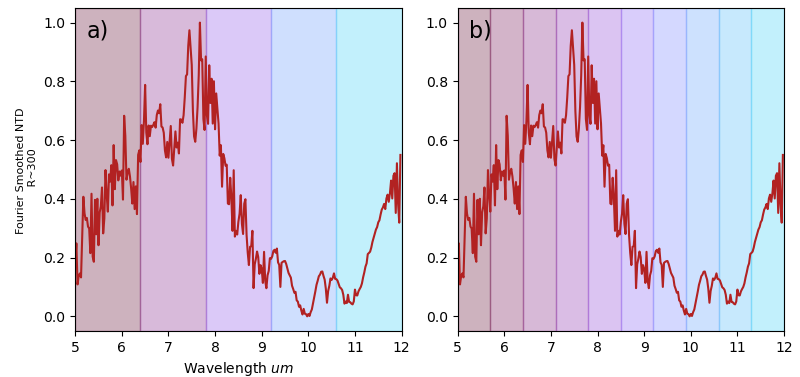}
\caption{The R$\sim$300 spectrum of Archean Earth around AD Leo with a) 5-part division and b) 10-part division.}
 \label{fig:FFTbands}
\end{figure}

We begin by dividing each spectrum into bands, under which we will integrate to create our primary metric. We use two methods of dividing our data for analysis: a 5-part division and a 10-part division. The 5-part division splits the spectrum into five 1.4$\mu$m increments from 5 to 12 $\mu$m. The 10-part division uses the same range but halves the width of each band to 0.7$\mu$m.  Figure \ref{fig:FFTbands} shows a degraded Archean AD Leonis spectrum as in Figure \ref{fig:FFTDegrade}c, with these two band separations highlighted by shaded regions. These two cases will be assessed based on their visual separation of the data and effectiveness at facilitating algorithmic clustering in Section \ref{sec:cluster}. 

With the bands set, we calculate the areas under the curves using the numerical Trapezoid method  within these partitions for the resolutions described in Figure \ref{fig:FFTDegrade}.

\subsubsection{Automatic Clustering Algorithm}
\label{sec:cluster}
Typical clustering algorithms, such as the commonly used ‘k-means’ method, function via the reduction of the sum of the squared error. Beginning with a number of user-specified points, the algorithms will compute the centroid of each cluster and successively move the points to that centroid, until new centroid computations do not change the position of the point. That is: when the centroids converge. This method is highly dependent on the initial choice of points and tend to bias towards spherical distributions of data, despite techniques to mitigate such errors \citep{Hennig2016}.

In this study, we will instead be using HDBSCAN (Hierarchical Density-Based Spatial Clustering of Applications with Noise), which is a different kind of clustering algorithm that operates off of the density of points within the target region. Each point is evaluated such that it is density-reachable or density-connected to at least one other point in its neighbourhood. Density-reachability is defined in \citet{EsterDBSCAN1996} as a point being within a radius of another point, such that this radius contains a minimum defined number of points (which determined the minimum density of the cluster). Density-connectivity is defined as a point \textbf{p} being reachable from point \textbf{q} via a point \textbf{o}, where \textbf{o} is density-reachable from both \textbf{p} and \textbf{q} even if \textbf{p} and \textbf{q} are not density-reachable from each other. This concept can be extended so that multiple points can be used, as long as the same density is considered to define the radii, to avoid crossing low-density regions separating the clusters.

\begin{figure}
 \noindent\includegraphics[width=\columnwidth]{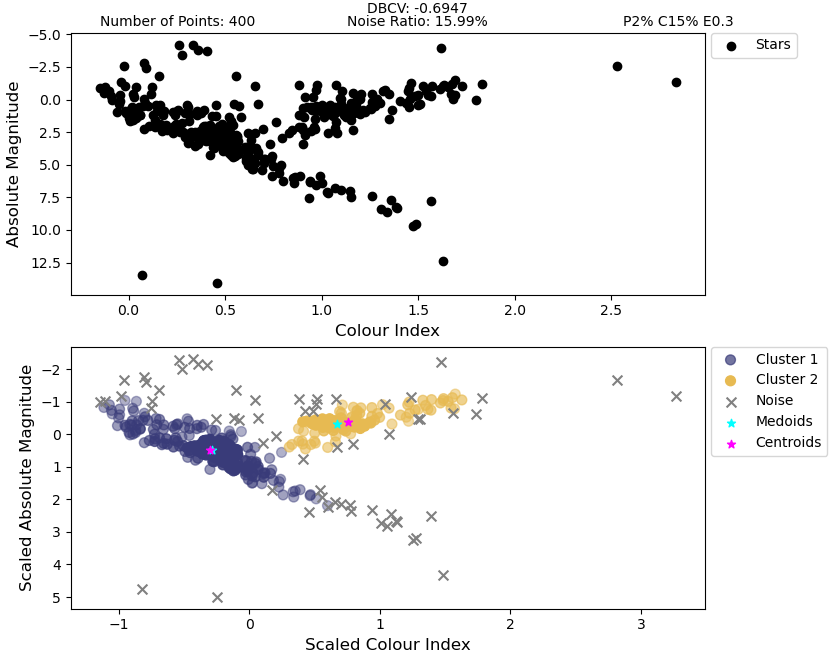}
\caption{Original (top panel) and HDBSCAN cluster analysis (bottom panel) of the Hertzsprung-Russel Diagram, the core of the Main Sequence auto-identified by HDBSCAN is highlighted in the purple Cluster 1, while the core of the Red Giant branch is highlighted in the yellow Cluster 2. The centroid and medoids of these clusters are shown in magenta and cyan stars, indicating the direction of the central “spine” of the cluster. Values detected by the algorithm to be "noise" have been marked with grey crosses  Example adapted from the discussion in \citet{CarollOstlie2007}}
 \label{fig:HRTest}
\end{figure}

Practically, this manifests as the algorithm requiring three input parameters: the minimum number of core points to form a cluster, the minimum number of points a cluster must have to be retained as one, and an “epsilon” value indicating how close two clusters can be before they are merged. These values are marked as P, C, and E respectively on Figure \ref{fig:HRTest}, Figure \ref{fig:Matrix} and Figure \ref{fig:DBSGrouped}, and are defined as a percentage value of the total number of points in the dataset.

HDBSCAN identifies circular and non-circular clusters while accounting for noise and outliers \citep{EsterDBSCAN1996}. Allowing for non-circular patterns is important because it will allow us to find linear patterns in scatter-plot data, and discern how it could potentially correlate to known quantities, thus uncovering links between spectrum and atmosphere. These algorithms do, however, require data preprocessed to remove variance and normalised around a zero mean. This scaling alters the calculated values of the areas beneath the curve but does not affect the distribution of the data relative to itself, which is the quantity we are concerned with in this study.

An example of HDBSCAN in use is shown in Figure \ref{fig:HRTest} using a Hertzsprung-Russell Diagram produced using the Hipparcos-Yale-Gliese stellar population data as a typical sample of astronomical data and distribution \citep{NASHTycho2024}. A k-means clustering algorithm would not find the linear relationship of the main sequence stars, whereas HDBSCAN (as found within the scikit.learn python toolkit) does, since it operates off of density rather than centroids.

Figure \ref{fig:HRTest} also shows valuation metrics used in this paper. DBCV is the Density Based Cluster Valuation detailed by \cite{Moulavi2014}. It is a relative cluster validation index formulated specifically to handle density-based clusters of arbitrary shape and spread. Traditionally, cluster validation is done through either comparing clustering results to the original labels of the data (inappropriate for unknown datasets such as the ones used in this paper) or assessing the Silhouette Coefficient. Silhouette Coefficient is used for spherical data clusters, which HDBSCAN does not necessarily produce.

DBCV is defined as:
\begin{equation}
  DBCV(C)=\sum_{i=1}^{i=l} \frac{|C_i|}{|O|}V_C(C_i)
  \label{eq:DBCV}
\end{equation}
where $C_i$ represents a cluster of index $i$, $O$ the total number of objects, together representing the relative size of the cluster, forming a weighted average in the sum. $V_C$ is validity of the cluster, based on its shape, as defined by,
\begin{equation}
  V_C(C_i)=\frac{\min_{1\leq j\leq l; j\neq i}(DSPC(C_i,C_j)-DSC(C_i))}{\max(\min_{1\leq j\leq l; j\neq i}(DSPC(C_i,C_j), DSC(C_i))}
  \label{eq:Validity}
\end{equation}

DSPC is the Density Separation of a Pair of Clusters, or the minimum distance between two points in different clusters. DSC is the Density Sparseness of a Cluster, is a measure of the compactness of a cluster, or the separation between points within a given cluster \citep{Moulavi2014}. Table \ref{tab:DBCV} summarises how DSC and DSPC contribute to DBCV.

\begin{table}
 \caption{The limits of DBCV clustering, showing how the DBCV value describes the shape of the clusters. No value is objectively “better” than another, rather, users must assess which shape is most appropriate for the data.}
 \centering
 \begin{tabular}{c | l}
 \hline
  DBCV Limit  & Shape  \\
 \hline
   $-1$  & DSPC$<$DSC, clusters are highly compact   \\
   $0$  & DSPC = DSC, clusters are balanced   \\
   $1$  & DSPC$>$DSC, clusters are spaced out   \\
 \hline
 \end{tabular}
 \label{tab:DBCV}
 \end{table}

The Noise Ratio is the percentage of the points labelled as noise by the algorithm. HDBSCAN also outputs the medoids and centroids of the clusters identified. Throughout this study we will be attempting to combine these values, along with visual assessment, to produce the best clusters possible. Minimising Noise Ratio ensures we take into account all of our data, and aiming for DBCV to be as close to zero as possible ensures we have balanced clusters.

\section{Results}
\subsection{Feature Separation}
\label{sec:ResFeat}
\begin{figure*} 
 \centering
 \includegraphics[width=\textwidth]{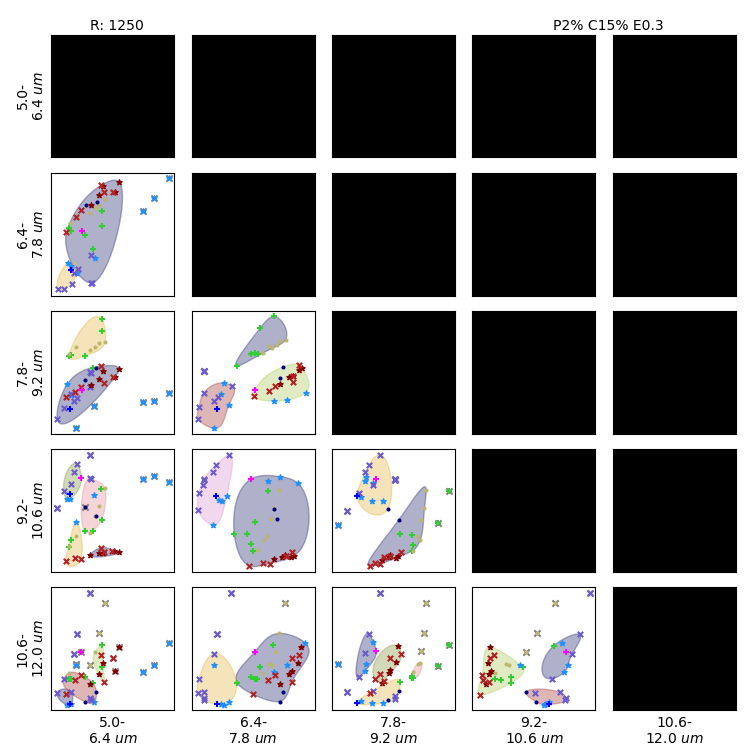}
  \caption{The 5-band division grid showing the area enclosed by the specified bands for all R$\sim$1250 planet spectra in this study. Data point colours and styles are as described in Table \ref{tab:Symbols}, with the intentional addition of an unknown sample (marked as a magenta cross). Shading represents the clusters as automatically determined by HDBSCAN.}
 \label{fig:Matrix}
\end{figure*}

We begin our analysis by comparing the integrated bands to each other. Using the R$\sim$1250 spectra with the 5-band division (similar to \ref{fig:FFTbands}a), the area enclosed in each band may be compared to that of the 4 others, resulting in a 5 x 5 matrix of cross-plots with 10 unique combinations, which are shown in Figure \ref{fig:Matrix}.

We can immediately observe the appearance of distinct clusters within the scatter plot data, and HDBSCAN also confirms the validity of these groupings. Because the HDBSCAN cluster analysis (and associated shading) is separate from our choices of colour and shape in Figure \ref{fig:Matrix}, we can confirm the clusters are not the result of pareidolic human error. Not all the cells of the matrix are useful, but others appear to separate clusters of planets which correlate with our prior knowledge of the planetary subtypes (here, the colours as defined in Table \ref{tab:Symbols}).

The method was repeated for the R$\sim$300 spectra of all samples with 10-part division, which resulted in 45 unique scatter plots and data clustering attempts and is displayed as Figure \ref{fig:ApB10}. While the specific values/cells are different in each case, both the 5 and 10 bin width divisions broadly contain multiple cells in which the auto-detected clustering appears to correlate with prior knowledge.

It is intriguing that this simple methodology separates specific planetary subtypes. To give an example, the ``Archean'' planets (red Xs and stars) are distinct from the ``Dry O$_2$'' planets (purple Xs) in the 9.2 -10.6µm vs 6.4-7.8µm cell in Figure \ref{fig:Matrix}, both ``by eye'' and as processed by HDBSCAN. This motivates additional study with agnostic feature detection and provides hope that additional planetary features might be extracted by this technique.

\subsection{Correlations of agnostic clusters with prior data}
\label{sec:ResCorr}
We can use cells in which the clusters identified by HDBSCAN are strongly separated in order to attempt to discern correlation and patterns. When a given planetary subtype correlates with the clustering observed in the cell, we may record it as a potential determinant of the separation. In addition, we can examine if the auto-detected clusters correlate with any of the planetary modeling data \citep{Arney_2017, Lincowski_2018, Meadows2018, Segura2005, Schwieterman_2016, Brune2023} as tabulated in Appendix \ref{sec:Plan}. We re-iterate that the position of the given planet in any one cluster plot is determined solely by the agnostic numerical integration of the spectrum. Overlaying (for example) the known mixing ratio values is thus independent of the creation of the separation matrices, and discovering a correlation here indicates that the species affect the spectrum of the light, either directly or through interaction with other molecules.

Within this dataset, no correlations between prior knowledge and auto-detected clusters were found when looking at planetary surface temperature, surface pressure, semi-major axis or instellation. However, we were able to identify surprisingly strong correlations to values (or ranges) of individual atmospheric components.

\begin{figure*}
\centering
\includegraphics[width=\textwidth]{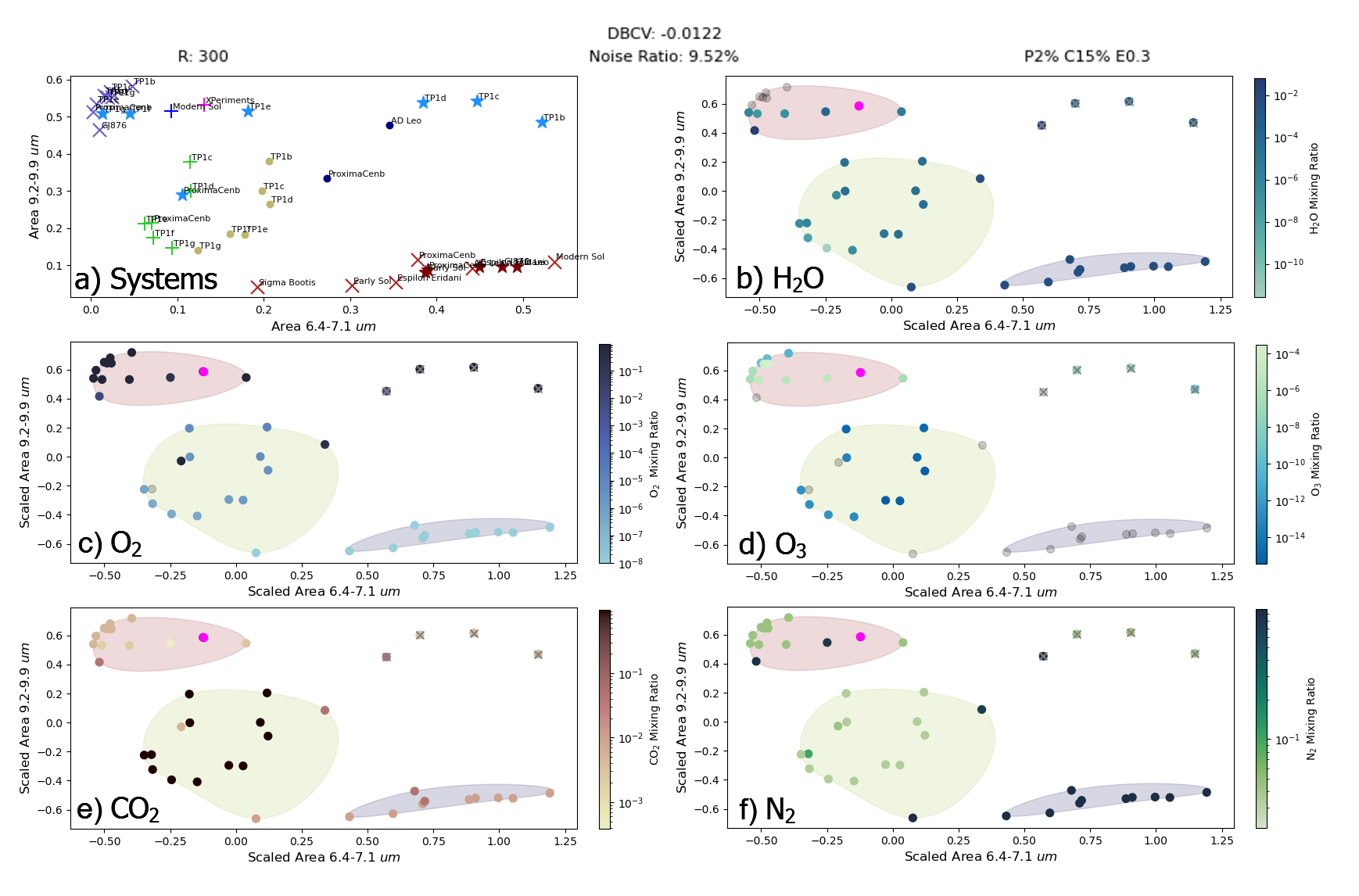}
\caption{a) cell F3
from Figure \ref{fig:ApB10} with symbols/colours from Table \ref{tab:Symbols}, b-f) HDBSCAN cluster analysis results (shaded regions), along with H$_2$O, O$_2$, O$_3$, CO$_2$ and N$_2$ mixing ratios presented as colour bars to the right of each subpanel. The magenta cross in panel a and magenta point in panels b-f represents the unknown sample.}
\label{fig:DBSGrouped}
\end{figure*}

Seven species commonly found within the planetary atmospheres of our models were chosen for analysis and their mixing ratios were recorded at a representative pressure of 1 bar (10$^{5}$ Pa). The mixing ratios for CO$_2$, CO, O$_2$, O$_3$, H$_2$O, N$_2$ and CH$_4$ are tabulated in Appendix \ref{sec:Plan}. Where data was neither available nor calculable from available information, the mixing ratio was noted as zero and represented by a grey circular data point in figures. Figure \ref{fig:DBSGrouped}a re-displays cell F3 of Figure \ref{fig:ApB10}, while \ref{fig:DBSGrouped}b-f show the planets coloured by the mixing ratios for five of these gases, with the auto-detected clusters showing as shaded regions.

Figure \ref{fig:DBSGrouped}b-f displays distinct clustering highlighting correlations with more than one species and implying that all of them contribute to the formation of the feature being extracted in these bands. We are thus reading not only absorption and emission features of individual molecules, but interference between them. Despite that, some patterns emerge which potentially isolate specific mixing ratios. 

\section{Discussion}

\subsection{Resolution}

During this investigation, it was found that while decreasing the resolution from full to approximately R$\sim$300 did affect the overall shape of the clusters, it did not significantly impact their ability to seperate unique clusters of data. This ability to seperate degrades quickly below R$\sim$300, and so does not provide useful clusters for our simulated MIRI-LRS dataset (R$\sim$30). Regardless of resolution adopted, the method did not constrain spectral type, instellation or surface pressure. Information about spectral type or instellation that may have been contained in the raw data would likely be degraded by our normalization procedure. Similarly, normalization degrades information about height of atmospheric radiation contained within the spectrum, although this is also somewhat degenerate with modeling assumptions.  It remains theoretically possible that other effects of pressure at the radiating layer (e.g. pressure broadening of lines) could be constrained through MASC, although we did not identify any evidence of this in this dataset.

\begin{figure}
 \includegraphics[width=\columnwidth]{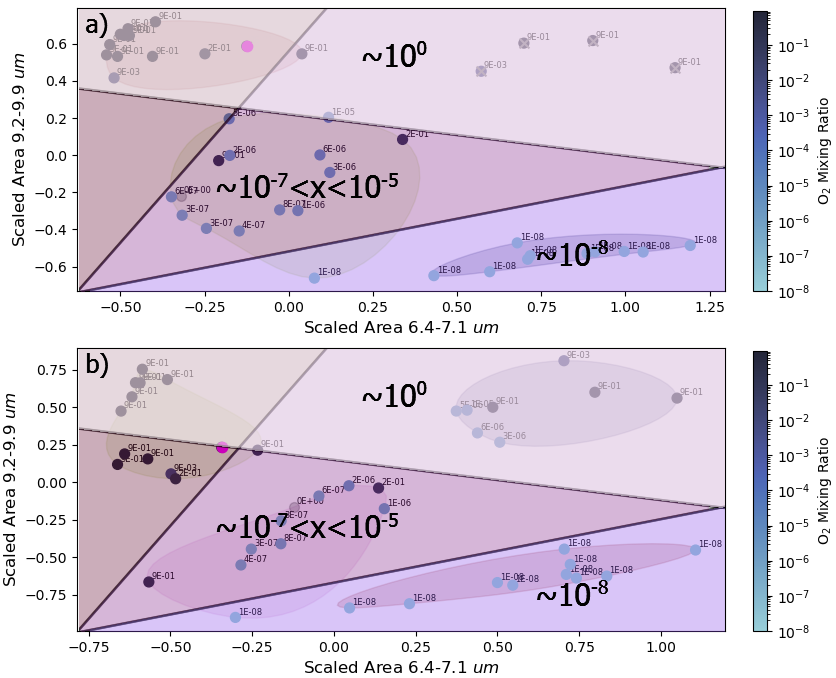}
  \caption{ Figure \ref{fig:DBSGrouped}c (Cell F3 from the 10-part analysis), showing overlaid O$_2$ mixing ratio (colourbar at right) with indications of four clustered regions that correlate with ranges in O$_2$ mixing ratio with a) R$\sim$300 data and b) Full resolution data.}
 \label{fig:Regions}
\end{figure}

Figure \ref{fig:Regions} illustrates the effect of resolution on our ability to provide mixing ratio constraints from one particular cell, while Table \ref{tab:Ranges} details some potential mixing ratio discriminants again arising solely from this one particular cell. We do not seek to argue that the two individual bands of the spectrum constrain these mixing ratios to the degree shown, but rather highlight the surprising potential constraints from one cell of the analysis.  While we are unable to explicitly describe the uncertainty associated with our predictions, our confidence in them would certainly grow if more than cells provided similar clear constraints.  We see this as an extrapolation of a common technique of molecular spectroscopy  -  confidence in line identification increases when more than one line is detected.

With the VPLSE model dataset reduced to R$\sim$300 , the concentrations form loose but distinct gradients in three bands on the scaled y-axis (Figure \ref{fig:Regions}a). The same analysis performed using the spectra with the full resolution from the VPLSE is shown as Figure \ref{fig:Regions}b. The analysis was also performed on the R$\sim$1250 data, but we found that the differences between it and the R$\sim$300 data were negligible. Therefore, the full resolution is shown to better display the effect of resolution on the cluster distribution.
The higher resolution data produces a similar effect, but with three triangular regions each displaying a different gradient, the lowermost of which generally being remarkably uniform. The central section typically will display either one or two gradient directions. In the case of ozone (not shown), there is no upper portion and the remaining non-uniform segment has a triple gradient vector. While N$_2$ does somewhat separate based on concentration, the effectiveness is low compared to other species. Table \ref{tab:Ranges} does not attempt to capture these various diagonals, but simply illustrates the value of one single measure (the normalized integrated area underneath the spectrum from 9.2-9.9 $\mu$m) in providing potential constraints on atmospheric mixing ratios.

\begin{table}
\caption{The approximate range of mixing ratios constrained in the clusters identified by HDBSCAN) and shown in Figure \ref{fig:Regions}a for O$_2$, using only the vertical axis.}
 \centering
 \begin{tabular}{ l || c c c }
\hline 
   & & Range (vertical axis) &\\
\cline{2-4}
   Species & -1:-0.5 & -0.5:0.5 & 0.5:1 \\
   & (Lower Region) & (Central Region) & (Upper Region)  \\
 \hline
   H$_2O$  & $10^{-3}$ & $10^{-8} \rightarrow 10^{-2}$ & $0 \rightarrow 10^{-4}$ \\
   O$_2$  & $10^{-8}$ & $10^{-7} \rightarrow 10^{-5}$ & $10^{-3} \rightarrow 10^{-1}$\\
   O$_3$  & 0  & $10^{-13} \rightarrow 0$ & $10^{-5} \rightarrow 10^{-8}$  \\
   CO$_2$  & $10^{-2}$  & $10^{0} \rightarrow 10^{-2}$ & $10^{-4} \rightarrow 10^{-2} $\\
   N$_2$  & $10^{0}$ & $10^{-2} \rightarrow 10^{-1}$ & $10^{-2}$ \\
 \hline
 \end{tabular}
 \label{tab:Ranges}
 \end{table}

\subsection{Ratios of Ratios}
\label{sec:ResRatios}
No single component can be a determinant for the full characterisation of its parent atmosphere. The reality of geochemical and atmospheric processes is that molecules interact with each other and that is what gives rise to surface conditions, rather than the individual presence of a given species. To reflect that, MASC must look at multiple components at once in our study of spectra. Combining bands that produced strong separation within individual mixing ratio plots (such as the one in Figure \ref{fig:DBSGrouped}b-f) provided a significant enhancement in the method's separation capabilities. 

\begin{figure}
 \includegraphics[width=\columnwidth]{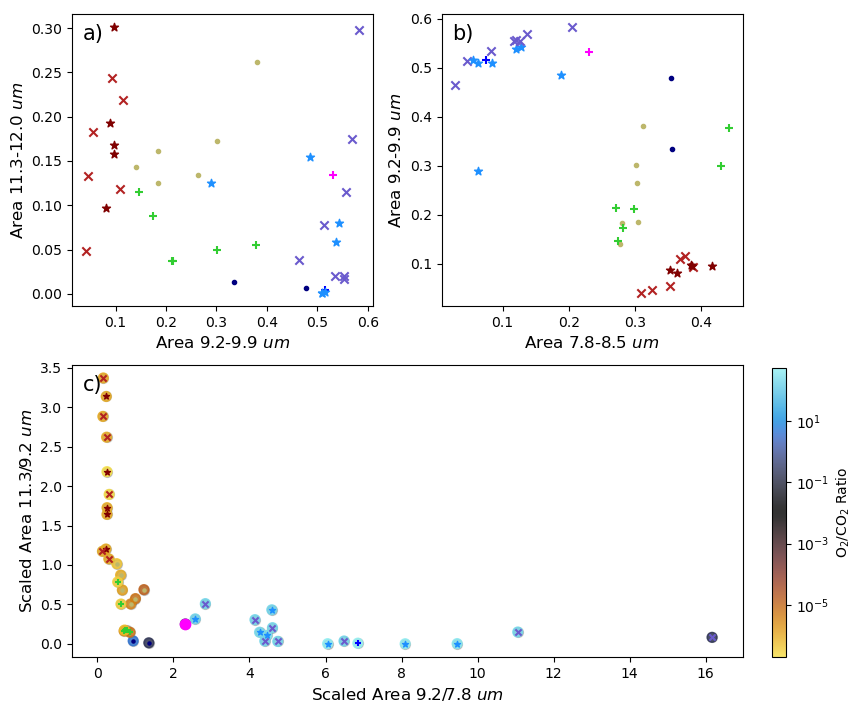}
 \caption{ The a) J6 and b) G4 cells from the 10x10 matrix with the signal reduced to R$\sim$300. The colours in panels a and b refer to the global legend in Table \ref{tab:Symbols}. c) The ratio of the cells from a) and b),  the colours correspond to the O$_2$/CO$_2$ ratio from the colour bar at right.}  
 \label{fig:Ratio}
\end{figure}

The cells chosen for trial in this method were chosen using the valuation metrics from Section \ref{sec:cluster}. We look for cells with a negative DBCV score close to zero (indicating clearly separated but not isolated clusters) and a Noise Ratio below 10$\%$. Among these, we prioritise rows and columns that feature multiple cells that match these criteria, indicating they produce strong and clear separation.
Thus, we arrive to the conclusion that columns 3 and 4 are strong candidates, as well as rows G and J. Through the mirroring effect of the diagonal, we add columns 6 and 9, and rows C and D. We then individually assess the quality of each remaining candidate cell, taking care to discard cells with overlapping or small ``spurious'' clusters that artificially lower the noise ratio.
Figure \ref{fig:Ratio} shows our strongest result through cells G4 and J6, demonstrating that the combination of two bands producing strong separation correlates extremely well with the known values of the O$_2$/CO$_2$ ratio.

Reduction in resolution causes negligible change to the pattern observed in Figure \ref{fig:Ratio}c, which shows a very clear separation and progression between O$_2$-rich and CO$_2$-rich atmospheres. This demonstrates that the MASC method is able to discern between these two types of planetary atmospheres from agnostic integration of the spectrum alone, even with relatively poor spectral resolution (R$\sim$300 compared to our own reduced MIRI MRS’s R$\sim$1250).

\subsection{Unknown Planet Identification}
\label{sec:ResID}
Now that we have demonstrate the manner in which agnostic metrics form clusters that correlate to planetary subtypes and mixing ratios, we can test the MASC method by attempting to identify an unknown sample. Our sample is a spectrum obtained from VPLSE but not considered in the previously described cluster or correlation analyses. With its planetary characteristics obscured, it can be analysed as an unknown variable before uncovering its ``true identity'' to confirm the hypothesis. In Figure \ref{fig:DBSGrouped}, Figure \ref{fig:Regions} and Figure \ref{fig:Ratio} this ``unknown" sample is magenta, and it also appears as a magenta cross in Figure \ref{fig:Matrix}.

In order to begin identification, we can refer to Table \ref{tab:Ranges} in order to give us a first idea of the type of planet we are dealing with. We note it is in the upper region of the plots in Figure \ref{fig:DBSGrouped}. Based on the other planets in its cluster, we can hypothesise the planet has relative concentrations of H$_2$O below $10^{-4}$, O$_2$ between $10^{-3}$ and 1, CO$_2$ between $10^{-4}$ and $10^{-2}$, and O$_3$ concentration between $10^{-5}$ and $10^{-8}$. When these values are observed as a ratio in Figure \ref{fig:Ratio}, we can observe it lies with the oxic atmospheres along the x-axis and among planets with an O$_2$/CO$_2$ ratio on the order of $10^{0}$, implying a near-equal proportion of O$_2$ and CO$_2$ in the atmosphere.
Thus, our final hypothesis to the identity of our unknown sample is a dry O$_2$ atmosphere similar to the rest of this subset. We verify this sample against its source \citep{Meadows2018}
and discover it is a desiccated O$_2$ model of Proxima Centauri b, containing CO$_2$, O$_2$ and CO (“FP Earth” in the VPLSE) and an ozone layer. This confirms part of our hypothesis, and also supports that this band-based separation functions as a method of resolving atmospheric subtypes at a distance. We underestimated the CO$_2$ mixing ratio, but note that our estimation was based on a single cell.  While additional work with larger datasets and more unknowns would be needed to more confidently make planetary identifications, the potential usefulness of the MASC method for combining information from multiple bands of spectra to constrain planetary characteristics appears robust. While MASC was demonstrated here for the far infrared and for transit transmission spectra, it should be feasible for reflected light spectra in the UV/visible/near IR, which will be the focus of a follow up effort.

\section{Conclusion}
In summary, using a dataset of 42 synthetic transit transmission spectra sourced from the Virtual Planetary Laboratory Spectral Explorer and adapted into a range matching that of JWST’s MIRI instrument, we examined the use of the normalized area enclosed by the curve in various wavelength bands as a method to characterise atmospheres. Spectra were reduced in resolution in order to simulate telescope functions and noise, normalised and split into 5 and 10 unique bands, and distribution of the enclosed areas were then assessed in a scatter plot. Unique pairs were identified using the HDBSCAN clustering algorithm.
A resolving power of R$\sim$300 was identified as the minimum resolving power necessary to produce useful results using automated cluster analysis.

Some of the automatically identified groupings further correlated to various known planet characteristics, such as planetary type, range of concentration of certain atmospheric components, and their ratios. We also analyzed an unknown sample using the Molecule Agnostic Spectral Clustering method developed here. We show that even with this small initial set at low resolution, MASC was sufficient to enable very strong constraints in the identification of an unknown planet.

We have thus demonstrated that it is possible to characterise exoplanet atmosphere type from agnostic interrogation of their transit spectra, and have extracted additional constraints from previously published data. In future work, we aim to refine our findings through an expanded dataset of synthetic planetary spectra to gain a more precise understanding of the constraints uncovered by this band-clustering method, investigate the feasibility of MASC in different wavelength regimes, as well as investigating its application to real-world data.

\section*{Acknowledgements}
We would like to thank Andrew Lincowski and Amber Young for kindly agreeing to share atmospheric data from their own work, and Victoria Meadows for her assistance with the Virtual Planetary Laboratory Spectral Explorer. 

This research has made use of the NASA Exoplanet Archive, which is operated by the California Institute of Technology, under contract with the National Aeronautics and Space Administration under the Exoplanet Exploration Program.





\bibliographystyle{mnras}
\bibliography{HicSunt.bib}




\appendix
\section{Additional Figures}
\label{sec:XTFigs}

\begin{figure*}
 \includegraphics[width=\textwidth]{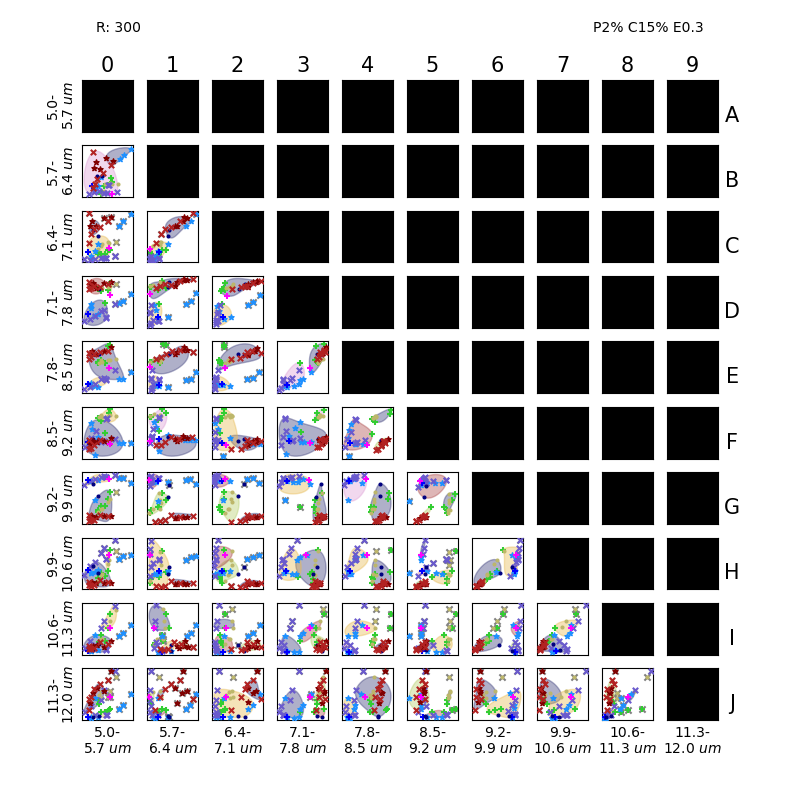}
 \caption{The 10-part grid showing the area enclosed by the bands for all planets in this study. Data point colours and styles are as described in \ref{tab:Symbols}, with the intentional addition of an unknown sample (marked as a magenta cross). Shading represents the clusters as automatically determined by HDBSCAN. The signal was reduced to R$\sim$300.}
 \label{fig:ApB10}
\end{figure*}

Figure \ref{fig:ApB10} shows an matrix of area values compared over 10 bands. This figure was generated using R$\sim$300 spectra.

\begin{figure*}
 \includegraphics[width=\textwidth]{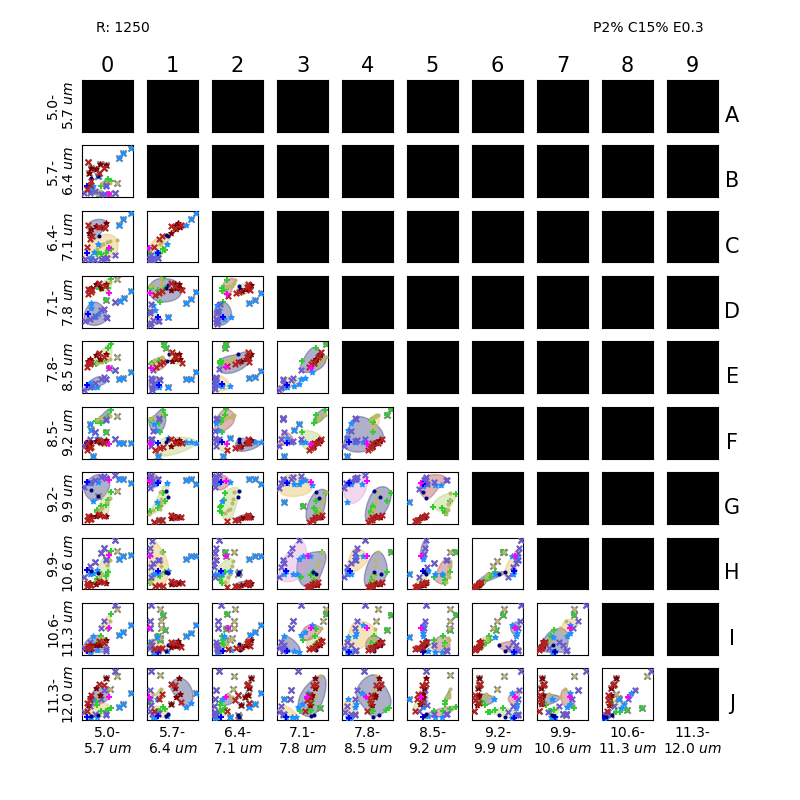}
 \caption{The 10-part grid showing the area enclosed by the bands for all planets in this study. Data point colours and styles are as described in \ref{tab:Symbols}, with the intentional addition of an unknown sample (marked as a magenta cross). Shading represents the clusters as automatically determined by HDBSCAN. The signal was reduced to R$\sim$1250.}
 \label{fig:ApB10.2}
\end{figure*}

Figure \ref{fig:ApB10.2} shows an matrix of area values compared over 10 bands. This figure was generated using R$\sim$1250 spectra.

\clearpage
\begin{landscape}
\section{Planet Data}
\label{sec:Plan}

The following (truncated) table shows the values for all measured species as well as key planetary properties for all 42 planets, ordered per host star.\\

    \begin{tabular}{lccccccccc}
Star Name&AD Leo&&&Modern Sol&&Archean Sol&&Epsilon Eridani&\\
Spectral Type&M&M&M&G&G&G&G&K2&K2\\
\hline
Atmosphere&Archean&Arch.Haze&E-type&Earth&Archean&Archean&Arch.Haze&Archean&Arch.Haze\\
T (K)&310&317&298&298&272&298&299&297&282\\
Flux ($S_{earth}$)&0.8&0.8&2&1&0.8&2&0.8&0.8&0.8\\
SemMaj (mAU)&1000&1000&1000&1000&1000&1000&1000&1000&1000\\
R ($R_{earth}$)&1&1&1&1&1&1&1&1&1\\
Pressure(bar)&1&1&&1&1&1&1&1&1\\
\hline
O$_2$&1.00E-08&1.00E-08&8.72E-03&2.10E-01&1.00E-08&1.00E-08&1.00E-08&1.00E-08&1.00E-08\\
H$_2O$&3.89E-03&3.89E-03&1.00E-02&1.00E-04&3.89E-03&3.89E-03&3.89E-03&3.89E-03&3.89E-03\\
CO&1.78E-04&1.78E-04&1.05E-02&0.00E+00&1.78E-04&1.78E-04&1.78E-04&1.78E-04&1.78E-04\\
CO$_2$&1.01E-02&1.01E-02&4.85E-02&3.85E-04&1.01E-02&1.01E-02&1.01E-02&1.01E-02&1.01E-02\\
O$_3$&0.00E+00&0.00E+00&0.00E+00&1.00E-05&0.00E+00&0.00E+00&0.00E+00&0.00E+00&0.00E+00\\
N$_2$&0&0&8.97E-01&7.81E-01&0.00E+00&0&0.00E+00&0&0.00E+00\\
CH$_4$&2.02E-03&2.02E-03&4.37E-04&1.82E-06&2.02E-03&2.02E-03&2.02E-03&2.02E-03&2.02E-03\\
\hline
\\
\hline
    \end{tabular}\\
    \begin{tabular}{lccccccccc}
Proxima Cen b&&&&&&&Sigma Bootis&GJ876&\\
M&M&M&M&M&M&M&F&M&M\\
Archean&E-type&WetO$_2$&DryO$_2$&FPEarth&Arch.Haze&V-type&Archean&DryO$_2$&Arch.Haze\\
278&273&318&256&254&277&379&277&290&301\\
0.66&0.66&0.66&0.66&0.66&0.66&0.66&0.8&1&0.8\\
48.5&48.5&48.5&48.5&48.5&48.5&48.5&1000&120&1000\\
1.074&1.074&1.074&1.074&1.074&1.074&1.074&1&1&1\\
1&1&10&10&1&1&10&1&&1\\
0.00E+00&2.06E-01&9.24E-01&9.35E-01&2.08E+00&0.00E+00&0&1.00E-08&8.72E-03&1.00E-08\\
5.80E-03&4.36E-03&3.35E-07&1.00E-06&0&6.06E-03&1.00E-06&3.89E-03&6.42E-02&3.89E-03\\
1.37E-08&1.93E-04&6.68E-08&5.99E-08&3.12E-01&1.40E-08&0&1.78E-04&1.05E-02&1.78E-04\\
5.02E-02&4.94E-02&4.95E-03&4.88E-03&5.10E-01&5.02E-02&8.95E-01&1.01E-02&4.85E-02&1.01E-02\\
0.00E+00&0.00E+00&0.00E+00&4.21E-07&5.88E-07&0.00E+00&0&0.00E+00&0.00E+00&0.00E+00\\
9.35E-01&7.32E-01&4.47E-02&4.45E-02&0&9.35E-01&1.01E-01&0&8.97E-01&0\\
1.02E-02&2.29E-03&0.00E+00&0.00E+00&0.00E+00&1.56E-02&0&2.02E-03&0.00E+00&2.02E-03\\
    \end{tabular}
\\
\\
\\
Full Table available on \href{https://github.com/archaeastra/ReadingBetweenRainbows/blob/main/MixingRatios.csv}{Github}
\end{landscape}

\bsp	
\label{lastpage}
\end{document}